\documentclass[12pt]{article}

\begin{document}

\begin{center}
{\bf Modified Gibbs's representation of rotation matrix} \\
\vspace{5mm}
 S. I. Kruglov \footnote{serguei.krouglov@utoronto.ca} and  V. Barzda \footnote{virgis.barzda@utoronto.ca}\\
\vspace{3mm}
Department of Physics, University of Toronto, \\60 St. Georges St.,
Toronto, ON M5S 1A7, Canada \\
Department of Chemical and Physical Sciences, \\
University of Toronto Mississauga,\\
3359 Mississauga Rd. N., Mississauga, ON L5L 1C6, Canada \\
Institute for Optical Sciences, University of Toronto, \\60 St. Georges St.,
Toronto, ON M5S 1A7, Canada
\vspace{5mm}
\end{center}

\begin{abstract}
A modified Gibbs's rotation matrix is derived and the connection with the Euler angles, quaternions, and Cayley$-$Klein parameters is established. As particular cases, the Rodrigues and Gibbs parameterizations of the rotation are obtained. The composition law of two rotations from the quaternion representation is presented showing a convenient expression for calculating the successive rotations.
\end{abstract}

\section{Introduction}

The aim of this investigation is to derive the most convenient form of the rotation matrix that would provide with the simplest equation describing combinations of several rotations, and to compare different rotation group parameterizations. The rotation group is defined as all rotation transformations around the origin of the 3D Euclidean space ($R^3$) and it is the Lie group. The group operations are smooth for the manifold of the structure. The rotation transformation is a linear transformation $x'_a=R_{ab}x_b$ (a,b=1,2,3 and, we imply a summation on repeated indexes, the Einstein summation), which leaves the length of the vectors invariant $\textbf{x}^{'2}=\textbf{x}^2$, and preserves the angles between the vectors (the $SO(3)$ group). Therefore, the rotation transformation preserves the dot product. The linear transformation with a reflection of all coordinates ($\textbf{x}\rightarrow -\textbf{x}$) is named an improper rotation. The composition of two rotations possesses the associative property and, therefore, it results in a different rotation that can be accomplished in one step. The relation $R_{ab}R_{cb}=\delta_{ac}$ (or $RR^T=I$, where $I$ is an identity matrix, $R^T$ is the transposed matrix, $R^T_{ab}=R_{ba}$, $\delta_{ac}$ is the Kronecker delta) follows from preservation of the vector length, i.e. $\textbf{x}^{'2}=\textbf{x}^2$. Matrices with the property $RR^T=I$ are called orthogonal matrices and corresponding transformations (the group $O(3)$) include proper and improper rotations. For orthogonal matrices, one obtains det$R=\pm 1$. If det$R=+1$ the transformations form the special orthogonal group $SO(3)$ (subgroup).
It is
the fact that the product of two orthogonal matrices with det$R=+1$ is still an orthogonal matrix with
det$R=+1$. But the transformations with det$R=-1$ are improper rotations that include the reflection. If we make two consequent improper rotations $R_1$ (det$R_1$=-1) and $R_2$ (det$R_2$=-1) then the resultant rotation is the proper rotation det$R_2R_1=1$ and it does not form a subgroup. For general two rotations $R_2R_1\neq R_1R_2$ which means that the rotation group is non-Abelian group, and therefore, it is important to follow the order for several consecutive rotations.

The Euler theorem states \cite{Euler}: ``\textit{the general displacement of a rigid body with one point fixed is a rotation about some axis}". According to Euler's rotation theorem \cite{Euler} every proper rotation (the fixed point is the origin) can be represented as a 2-dimensional single rotation about some axis by an angle $\alpha$ in the plane orthogonal to this axis. The axis of the rotation can be characterized by the unit vector $\hat{\textbf{n}}$ ($\hat{\textbf{n}}^2=1$) with two independent parameters, which remain unchanged by the rotation. If the angle of the rotation is zero then the axis is not uniquely defined. Thus, the rotation group possesses three independent parameters (three degrees of freedom) $\hat{\textbf{n}}$, $\alpha$.
One can use a non-normalized 3-vector $\textbf{v}$ (a rotation vector) with the length depending on the angle of rotation ($\alpha$). It should be mentioned that $\textbf{v}$ is not a real vector because for two successive rotations with vectors $\textbf{v}_1$ and $\textbf{v}_2$ the resultant rotation vector $\textbf{v}$ is not the sum of $\textbf{v}_1$ and $\textbf{v}_2$ ($\textbf{v}\neq\textbf{v}_1+\textbf{v}_2$). Therefore, the finite rotation is not a real vector, and we can consider the $\textbf{v}$ as a collection of three parameters $(v_1, v_2, v_3)$.  If $\textbf{v}=\tan(\alpha/2) \hat{\textbf{n}}$, one has the Gibbs representation \cite{Gibbs}, \cite{Gibbs1} and if $\textbf{v}=\alpha \hat{\textbf{n}}$, we come to the Rodrigues parameter \cite{Rodrigues}. In this paper, we use the rotation vector with the normalization $\textbf{v}=\sin(\alpha/2)\hat{\textbf{n}}$ (see also \cite{Euler2}, \cite{Gauss}, \cite{Fedorov}, \cite{Marsden}). Such representation has some advantage compared to Rodrigues and Gibbs representations, and thus can be successfully used in various applications. Another way to describe the rotation is to use three Euler angles \cite{Joshi}, \cite{Goldstein}, \cite{Landau}, quaternions \cite{Casanova}, \cite{Wertz}, \cite{Altmann}, \cite{Shuster}, \cite{Kruglov2} and Cayley$-$Klein parameters \cite{Cayley}, \cite{Goldstein}, \cite{Arfken}, \cite{Gel'fand}. It should be mentioned that all representations are suitable
explicit forms of the exponential map:
\[
S^1\times S^2\ni (\alpha,\hat{\textbf{n}})\mapsto \exp\{\alpha W(\hat{\textbf{n}})\}\in SO(3),~~~~W^T=-W,
\]
$S^2=SO(3)/S^1$, and the Euler theorem is by no means global.

The goal of this paper is to derive a convenient parameterization of the rotation group and to obtain the simplest composition law of two rotations within the modified Gibbs parametrization. This is important for cases when there are many rotations and the resultant rotation matrix has to be obtained.

The paper is organized as follows: In Sec. 2, we derive the general expression for the rotation matrix with non-normalized rotation vector. The connection of the rotation vector with Euler angles, quaternion, and Cayley$-$Klein parameters is presented in Sec. 3. In Sec. 4, we obtain the composition of two rotations from the quaternion rotation representation. This allows us to find the resultant rotation parameter in a simple manner for the case of many rotations. The Rodrigues and Gibbs representations of the parametrization are given in Sec. 5 as a particular cases, followed by conclusions of the article in Sec. 6.

We use the convention that Latin letters run as $1,2,3$ and Greek letters run as $1,2,3,4$.

\section{The rotation matrix}

The rotation matrix $R$ depends on three independent parameters and any antisymmetric matrix $A$ ($A_{ab}=-A_{ba}$) possesses tree independent parameters. Therefore, we can express the matrix $R$ via the matrix $A$. Arbitrary antisymmetric matrix $A$ obeys the minimal matrix equation as follows:
\begin{equation}
A^3=\frac{1}{2}\textrm{Tr}\left(A^2\right)A,
\label{1}
\end{equation}
where Tr$(A^2)$ is a trace of the matrix $A^2$. We may look for the expression of $R$ in the following form
\begin{equation}
R=c_0I+c_1A+c_2A^2,
\label{2}
\end{equation}
where $I$ is an identity matrix, and $c_0$, $c_1$, and $c_2$ are arbitrary coefficients. As $A^T=-A$, one arrives at the transposed matrix
\begin{equation}
R^T=c_0I-c_1A+c_2A^2.
\label{3}
\end{equation}
Then from the orthogonality condition $RR^T=1$ and Eqs. (1)-(3), we come to
\begin{equation}
c_0=1,~~~~c_1^2=2c_2+\frac{1}{2}c_2^2\textrm{Tr}\left(A^2\right).
\label{4}
\end{equation}
It is convenient to use the expression for the antisymmetric matrix $A$ as follows:
\begin{equation}
A_{ab}=\epsilon_{acb}a_c,
\label{5}
\end{equation}
where $\epsilon_{acb}$ is the Levi$-$Civita symbol, and the $a_c$ are three parameters. The matrix $A$ (Eq. (5)) possesses the attractive property: defining the vector $c_i=A_{ij}b_j$, where $\textbf{b}$ is an arbitrary vector, we obtain the cross product $\textbf{c}=\textbf{a}\times\textbf{b}$.
We will show later that $a_c$ is connected with a rotation vector. The vector $a_c$ can be found from the matrix elements of the matrix $A$:
\begin{equation}
a_c=\frac{1}{2}\epsilon_{acb}A_{ab}.
\label{6}
\end{equation}
Thus, there is one to one correspondence between the antisymmetric matrix $A$ and the vector $\textbf{a}=(a_1,a_2,a_3)$. From Eq. (5), one obtains Tr$A^2=-2\textbf{a}^2$, and from Eq. (4) the rotation matrix (2) becomes
\begin{equation}
R=I+\sqrt{c_2\left(2-c_2\textbf{a}^2\right)}A+c_2A^2.
\label{7}
\end{equation}
From Eq. (5), we find $(A^2)_{ij}=a_ia_j-\textbf{a}^2\delta_{ij}$. Let us introduce a new vector $\textbf{b}=\sqrt{c_2/2}\textbf{a}$. Then we obtain the rotation matrix (7) in the form (see also \cite{Fedorov}, \cite{Marsden})
\begin{equation}
R_{ij}=\left(1-2\textbf{b}^2\right)\delta_{ij}+2\sqrt{1-\textbf{b}^2}\epsilon_{ikj}b_k+2b_ib_j.
\label{8}
\end{equation}
Eq. (8) is the general expression for the rotation matrix via three parameters $\textbf{b}=(b_1,b_2,b_3)$. Thus, the coefficient $c_2$ is absorbed and we get to only three independent parameters. From  Eq. (8), one has the restriction $|\textbf{b}|\leq 1$. The $3\times 3$-matrix (8) is given by
\begin{equation}
R=\left(
  \begin{array}{ccc}
    1-2b_2^2-2b_3^2 & -2\sqrt{1-\textbf{b}^2}b_3+2b_1b_2 &2\sqrt{1-\textbf{b}^2}b_2+2b_1b_3 \\
    2\sqrt{1-\textbf{b}^2}b_3+2b_1b_2 & 1-2b_1^2-2b_3^2 & -2\sqrt{1-\textbf{b}^2}b_1+2b_2b_3 \\
    -2\sqrt{1-\textbf{b}^2}b_2+2b_1b_3 & 2\sqrt{1-\textbf{b}^2}b_1+2b_2b_3 & 1-2b_1^2-2b_2^2 \\
  \end{array}
\right),
\label{9}
\end{equation}
with the trace Tr$R=3-4\textbf{b}^2$. At $\textbf{b}=0$ the matrix (9) becomes the unit matrix $R=1$ showing the absence of rotation. One can verify that indeed the matrix (9) obeys the orthogonality condition $RR^T=I$. Calculating the determinant of the matrix (9), we obtain det$R$=1, i.e. the matrix (8) (or (9)) describes the proper rotations and therefore corresponds to the special orthogonal group $SO(3)$. If the orthogonal matrix $R$, which describes the proper rotation, is given, we can find the rotation vector $\textbf{b}$. Indeed, we find from Eq. (9) the relations as follows:
\begin{equation}
1+\textrm{Tr}R=4\left(1-\textbf{b}^2\right),~~~~\left(R-R^T\right)_{ij}=4\sqrt{1-\textbf{b}^2}\epsilon_{ikj}b_k.
\label{10}
\end{equation}
From Eq. (10), one obtains:
\begin{equation}
b_k=\frac{\left(R-R^T\right)_{ij}\epsilon_{ikj}}{4\sqrt{1+\textrm{Tr}R}}.
\label{11}
\end{equation}
Eq. (11) allows us to find the unique rotation vector $\textbf{b}$ for a given rotation matrix $R$. The only exception is the case $|\textbf{b}|=1$. It follows then from Eq. (10) that $R=R^T$ and $1+\textrm{Tr}R=0$. In this case we can not use Eq. (11), but from Eq. (8), one finds
\begin{equation}
R_{ij}=2b_ib_j-\delta_{ij}.
\label{12}
\end{equation}
To obtain the connection of the angle of rotation with the vector $\textbf{b}$, we consider rotating the vectors $\textbf{c}_1$, which is parallel to $\textbf{b}$ ($\textbf{c}_1\|\textbf{b}$), and $\textbf{c}_2$, which is perpendicular to $\textbf{b}$ ($\textbf{c}_2\perp\textbf{b}$), using the matrix (12). As a result, we obtain
\begin{equation}
R_{ij}\left(c_1\right)_j=\left(c_1\right)_i, ~~~~R_{ij}\left(c_2\right)_j=-\left(c_2\right)_i.
\label{13}
\end{equation}

It is seen, therefore, that the rotation matrix (12) with the unit rotation vector $|\textbf{b}|=1$ corresponds to the rotation with the $180^\circ$ angle. It follows from Eq. (8) that for any vector $\textbf{d}$ parallel to $\textbf{b}$ ($\textbf{d}\|\textbf{b}$), we have $R(\textbf{b})\textbf{d}=\textbf{d}$. Therefore the matrix $R$ rotates vectors about the vector $\textbf{b}$, i.e. the $\textbf{b}$ defines the axis of the rotation. To clear up the meaning of the $\textbf{b}$ vector length, let us consider the rotation by an angle $\alpha$ of the unit vector $\widehat{\textbf{c}}$, which is perpendicular to the vector $\textbf{b}$ ($\widehat{\textbf{c}}\perp \textbf{b}$). Then the cosine of the angle is given by the dot product and with the help of Eq. (8) we get:
\begin{equation}
\cos\alpha=\widehat{\textbf{c}}\cdot\widehat{\textbf{c}}'=\widehat{\textbf{c}}_iR_{ij}\widehat{\textbf{c}}_j=
1-2\textbf{b}^2.
\label{14}
\end{equation}
From Eq. (14) we obtain the expression for the length of the vector $\textbf{b}$:
\begin{equation}
|\textbf{b}|=\sin\frac{\alpha}{2}.
\label{15}
\end{equation}
It obeys the necessary condition $|\textbf{b}|\leq 1$. By introducing the unit vector $\widehat{\textbf{b}}$ ($\textbf{b}=\widehat{\textbf{b}}\sin\alpha/2$), Eq. (8) takes a simple form (see also \cite{Fedorov}, \cite{Marsden}):
\begin{equation}
R_{ij}=\cos\alpha\delta_{ij}+\sin\alpha\epsilon_{ikj}\widehat{b}_k
+2\sin^2\frac{\alpha}{2}\widehat{b}_i\widehat{b}_j.
\label{16}
\end{equation}
It is not difficult to obtain the resultant matrix $R$ from Eq. (16) for two successive rotations $R_1$ and $R_2$, $R=R_2R_1$. The rotation matrix in the form of Eq. (16) can be useful for various applications.

\section{The connection of the rotation vector with Euler angles, quaternions and Cayley$-$Klein parameters}

\subsection{Euler angles}

Any rotation can be represented as three successive rotations: a rotation by an angle $\varphi$ around the axis Oz ($R_\varphi$), then a rotation by an angle $\theta$ around the Ox axis ($R_\theta$), and a rotation by an angle $\psi$ around the axis Oz' ($R_\psi$) \cite{Gel'fand}. The rotation matrix $R_E=R_\psi R_\theta R_\varphi$ in terms of three Euler angles $\varphi$, $\psi$, and $\theta$ reads
\begin{equation}
R_E=\left(
    \begin{array}{ccc}
      \cos\varphi\cos\psi-\cos\theta\sin\varphi\sin\psi & -\cos\varphi\sin\psi-\cos\theta\sin\varphi\cos\psi & \sin\varphi\sin \theta\\
      \sin\varphi\cos\psi+\cos\theta\cos\varphi\sin\psi & -\sin\varphi\sin\psi+\cos\theta\cos\varphi\cos\psi & -\cos\varphi\sin\theta \\
     \sin\psi\sin \theta & \cos\psi\sin \theta & \cos\theta \\
    \end{array}
  \right)
.
\label{17}
\end{equation}
The range of the angles $\varphi$ and $\psi$ is $(0,2\pi)$, and the range of the angle $\theta$ is $(0,\pi)$. If $\theta =0$ the corresponding rotation is around Oz by the angle $\varphi+\psi$. The inverse transformation $R^{-1}$ is defined by the angles $\pi-\psi$, $\theta$ and $\pi-\varphi$. From Eq. (17), we obtain $1+$Tr$R_E=\left(1+\cos\theta\right)\left(1+\cos(\varphi+\psi)\right)$. One can find the rotation vector $\textbf{b}$ expressed via Euler angles $\varphi$, $\psi$, and $\theta$ from Eq. (11). We notice that the matrix $R-R^T$ is antisymmetric, and therefore, components of the rotation vector, found from Eq. (11), are given by
\begin{equation}
b_1=\frac{\left(R-R^T\right)_{32}}{4\sqrt{1+\textrm{Tr}R}},~~
b_2=\frac{\left(R-R^T\right)_{13}}{4\sqrt{1+\textrm{Tr}R}},~~
b_3=\frac{\left(R-R^T\right)_{21}}{4\sqrt{1+\textrm{Tr}R}}.
\label{18}
\end{equation}
After some calculations, from Eqs. (17), (18), we obtain the components of the rotation vector expressed via Euler angles:
\[
b_1=\sin\frac{\theta}{2}\cos\left(\frac{\varphi-\psi}{2}\right),~~
b_2=\sin\frac{\theta}{2}\sin\left(\frac{\varphi-\psi}{2}\right),
\]
\vspace{-8mm}
\begin{equation}
\label{19}
\end{equation}
\vspace{-8mm}
\[
b_3=\cos\frac{\theta}{2}\sin\left(\frac{\varphi+\psi}{2}\right).
\]
The angle of rotation about the vector $\textbf{b}$ may be obtained from Eq. (15). Evaluating the length of the vector $\textbf{b}$ from Eq. (19) and taking into account Eq. (15), we find
\begin{equation}
\cos\frac{\alpha}{2}=\cos\frac{\theta}{2}\cos\left(\frac{\varphi+\psi}{2}\right).
\label{20}
\end{equation}
Thus, knowing three Euler angles $\varphi$, $\psi$, and $\theta$, one can calculate from the Eqs. (19) and (20) the direction of rotation axis, $\textbf{b}$, and the angle of rotation. At the same time if we know a rotation vector $\textbf{b}$, the three Euler angles can be obtained from Eq. (19):
\[
\theta=2\arcsin\sqrt{b_1^2+b_2^2},~~~ \varphi=\arcsin\frac{b_2}{\sqrt{b_1^2+b_2^2}}+\arcsin\frac{b_3}{\sqrt{1-b_1^2-b_2^2}},
\]
\vspace{-8mm}
\begin{equation}
\label{21}
\end{equation}
\vspace{-8mm}
\[
\psi=-\arcsin\frac{b_2}{\sqrt{b_1^2+b_2^2}}+\arcsin\frac{b_3}{\sqrt{1-b_1^2-b_2^2}}.
\]
Thus, there is one to one correspondence between the rotation vector $\textbf{b}$ and three Euler angles.

\subsection{Quaternions}

Quaternions can be considered as a generalization (doubling) of the complex numbers \cite{Hamilton}. There are many applications of quaternions including a spacecraft altitude estimation \cite{Bar}, \cite{Markley}, \cite{Markley1}, investigation of the symmetry of the fields, and formulation of the relativistic wave equations \cite{Kruglov2}, \cite{Kruglov}, \cite{Kruglov1}. The quaternion algebra is defined by four basis elements $e_\mu =(e_k,e_4)$ with the
multiplication properties \cite{Casanova}:
\[
e_4^2=1,~~e_1^2=e_2^2=e_3^2=-1,~~e_1e_2=-e_2e_1=e_3,
\]
\vspace{-8mm}
\begin{equation}
\label{22}
\end{equation}
\vspace{-8mm}
\[
e_2e_3=-e_3e_2=e_1,~~e_3e_1=-e_1e_3=e_2,~~e_4e_m=e_me_4=e_m,
\]
where $m=1,$ $2,$ $3$, and $e_4=1$ is the unit element. The quaternion algebra, Eq. (22), can be constructed with the help of the Pauli matrices:
\begin{equation}
\sigma_0=\left(
           \begin{array}{cc}
             1 & 0 \\
             0 & 1 \\
           \end{array}
         \right),~
         \sigma_1=\left(
           \begin{array}{cc}
             0 & 1 \\
             1 & 0 \\
           \end{array}
         \right),~
         \sigma_2=\left(
           \begin{array}{cc}
             0 & -i \\
             i & 0 \\
           \end{array}
         \right),~
         \sigma_3=\left(
           \begin{array}{cc}
             1 & 0 \\
             0 & -1 \\
           \end{array}
         \right).
\label{23}
\end{equation}
The Pauli matrices (23) obey the relations as follows:
\[
\sigma _m\sigma _n=i\epsilon _{mnk}\sigma _k+\delta _{mn},~~\sigma_\mu \overline{\sigma }_\nu +\sigma_\nu \overline{\sigma}_\mu=-2\delta_{\mu \nu},
\]
\vspace{-8mm}
\begin{equation}
 \label{24}
\end{equation}
\vspace{-8mm}
\[
\sigma _\mu =\left( \sigma_k,\sigma_4\right) ,\hspace{0.3in}
\overline{\sigma}_\mu =\left( -\sigma_k,\sigma_4\right) ,
\]
where $\sigma_4=i\sigma_0$. By setting $e_4=\sigma_0$, $e_k=i\sigma_k$ and with the help of Eq. (24) we obtain the properties presented in Eq. (22). Let us consider the quaternion
\begin{equation}
q=b_\mu e_\mu =b_me_m+b_4e_4,  \label{25}
\end{equation}
with the rotation vector $\textbf{b}=\widehat{\textbf{b}}\sin(\alpha/2)$, and the scalar term $b_4=\cos(\alpha/2)$ (see also \cite{Rodrigues}, \cite{Euler2}, \cite{Gauss}). Four parameters $b_\mu$ can be called the Euler-Rodrigues parameters. Thus, the vector $\textbf{b}$ can be used as the matrix representation, Eqs. (8) and (9), as well as the quaternion, Eq. (25). But the quaternion representation of rotation, Eq. (25), requires introduction of the fourth component $b_4$. The operation of quaternion conjugation is defined as
\begin{equation}
\overline{q}=b_4e_4-b_me_m\equiv q_4-\mathbf{q}. \label{26}
\end{equation}
One can verify that the equalities $\overline{q_1+q_2}=\overline{q}_1+\overline{q}_2$, $\overline{
q_1q_2}=\overline{q}_2\overline{q}_1$ hold for two arbitrary
quaternions $q_1$ and $q_2$. The quaternion modulus $\mid
q\mid $ is defined by the relation $ \mid q\mid =\sqrt{q\overline{q}}=\sqrt{q_\mu ^2}$. The quaternion algebra includes the division, e.g. $q_1/q_2=q_1\overline{q}_2/ \mid q_2\mid^2$. For our rotational quaternion, Eq. (25), the modulus is equal to unity, $\mid q\mid=1$. Thus, unit quaternion represents the three-dimensional sphere $S^3$ ($b_\mu^2=1$) embedded in 4-dimensional Euclidean space $E^4$. The finite rotation transformations are given by \cite{Casanova}:
\begin{equation}
{\bf x}^{\prime }=q{\bf x}\overline{q},
\label{27}
\end{equation}
where ${\bf x}=x_me_m$ is the quaternion of the spatial coordinates or some three component vector, $q$ is the quaternion of the rotation group with the constraint $q\overline{q}=1$, and is given by Eq. (25). We note that $q$ and $(-q)$ represent the same rotation, i.e. there is a sign ambiguity. One may verify that the squared vector of coordinates, $x_m ^2$, is invariant under the transformations, Eq. (27). Indeed, $x_m^{\prime 2}={\bf x}^{\prime}\overline{{\bf x}}^{\prime}=q{\bf x}\overline{q}q\overline{{\bf x}}\overline{q}={\bf x}\overline{{\bf x}}=x_m ^2$, as $\overline{q}q=q\overline{q}=1$. Thus, the $3$-parameter transformations (Eq. (27)) and the rotational quaternion (Eq. (25)) belong to the special orthogonal group $SO(3)$. The rotational transformations in the form of Eq. (27) have some advantages compared to the form $x'_m=R_{mn}x_n$ because $R_{mn}$ are $3\times 3$-matrices but quaternions can be realized by $2\times 2$ Pauli matrices with complex elements. Therefore, it is easy to combine two individual rotations by the quaternion representation. It should be noted that the quaternion multiplication has inverse ordering ($q_1q_2$) compared to the matrix product ($R_2R_1$). We also note that quaternions $q$ and ($-q$) represent the same rotation. The quaternion representation of the rotation is very convenient because the quaternion varies continuously over $S^3$ when rotation angles change, and there are no jumps, which take place with some three-dimensional parameterizations.

\subsection{Cayley$-$Klein parameters}

With the help of Pauli matrices, Eq. (23), and the replacement $e_4=\sigma_0$, $e_k=i\sigma_k$, the quaternion, Eq. (25), can be represented as a matrix
\begin{equation}
Q=b_4\sigma_0+ib_m\sigma_m=\left(
                             \begin{array}{cc}
                               b_4+ib_3 & b_2+ib_1 \\
                               -b_2+ib_1 & b_4-ib_3\\
                             \end{array}
                           \right).
\label{28}
\end{equation}
Thus, the matrix $Q$ has a close relationship with quaternions. One can verify that det$Q=1$ and $QQ^+=Q^+Q=1$ ($Q^+$ is Hermitian conjugated matrix) as $b_\mu^2=1$, and the matrix $Q$ is the unitary matrix. The matrix $Q$ in Eq. (28) can be parameterized by Cayley$-$Klein parameters as follows (see, for example \cite{Goldstein}, \cite{Gel'fand}):
\begin{equation}
Q=\left(
        \begin{array}{cc}
         \alpha & \beta \\
         -\beta^\ast & \alpha^\ast\\
         \end{array}
         \right),
\label{29}
\end{equation}
where $\alpha^\ast$, $\beta^\ast$ are complex conjugated parameters, with the restriction $|\alpha|^2+|\beta|^2=1$, so that $QQ^+=Q^+Q=1$, and there are three independent degrees of freedom characterizing the rotation. Comparing Eq. (28) and Eq. (29), we obtain Cayley$-$Klein parameters expressed via the components of the rotation vector
\begin{equation}
\alpha=b_4+ib_3,~~~~\beta=b_2+ib_1.
  \label{30}
\end{equation}
Parameter $b_4$ can be expressed through the Euler angles as follows:
\begin{equation}
b_4=\cos\frac{\alpha}{2}=\cos\frac{\theta}{2}\cos\left(\frac{\varphi+\psi}{2}\right).
  \label{31}
\end{equation}
One can also represent Cayley$-$Klein parameters via the Euler angles by using Eqs. (19), (31)
\begin{equation}
\alpha=\cos\frac{\theta}{2}\exp\left(\frac{\psi+\varphi}{2}\right), ~~~~\beta=i\sin\frac{\theta}{2}\exp\left(\frac{\psi-\varphi}{2}\right).
  \label{32}
\end{equation}
If we introduce the matrix
\begin{equation}
X=x_m\sigma_m=\left(
        \begin{array}{cc}
         x_3 & x_1-ix_2\\
         x_1+ix_2 & -x_3\\
         \end{array}
         \right),  \label{33}
\end{equation}
then finite rotation transformations read
\begin{equation}
X^{\prime }=QXQ^+.  \label{34}
\end{equation}
We note that Hermitian conjugate matrix $Q^+$ corresponds to the conjugate quaternion (Eq. (26)). For two successive rotations with 2$\times$2 matrices $Q_1$ and $Q_2$, the resultant matrix is $Q_1Q_2$ (and corresponding quaternion $q_1q_2$), i.e. the order is opposite compared with the 3$\times$3 rotation matrices $R_1$ and $R_2$ for which the resultant matrix is $R_2R_1$. Expressing the components of the rotation vector via Cayley$-$Klein parameters from Eq. (30)
\begin{equation}
b_1=\frac{i\left(\beta^\ast-\beta\right)}{2},~~b_2=\frac{\left(\beta+\beta^\ast\right)}{2},~~
b_3=\frac{i\left(\alpha^\ast-\alpha\right)}{2},~~b_4=\frac{\left(\alpha+\alpha^\ast\right)}{2},
  \label{35}
\end{equation}
one obtains using Eq. (9) the rotation 3$\times$3 matrix
\begin{equation}
R=\left(
  \begin{array}{ccc}
   \frac{1}{2}\left(\alpha^2+ \alpha^{\ast 2}-\beta^2-\beta^{\ast 2}\right)& \frac{i}{2}\left(\alpha^2- \alpha^{\ast 2}-\beta^2+\beta^{\ast 2}\right) &\beta\alpha^\ast+\alpha\beta^\ast\\
   \frac{i}{2}\left(\alpha^{\ast 2}-\alpha^2-\beta^2+\beta^{\ast 2}\right)&  \frac{1}{2}\left(\alpha^2+ \alpha^{\ast 2}+\beta^2+\beta^{\ast 2}\right) & i\left(\beta\alpha^\ast-\alpha\beta^\ast\right)\\
   -\beta\alpha-\alpha^\ast\beta^\ast & i\left(\alpha^\ast\beta^\ast-\alpha\beta\right)&  \alpha\alpha^\ast- \beta\beta^\ast\\
  \end{array}
\right).
\label{36}
\end{equation}
Thus, for any unitary matrix $Q$ ($QQ^+=1$) (29) with the determinant $1$ ($\det Q=1$) there is a definite rotation matrix (Eq. (36)). It should be noted that if one makes the replacement $\alpha\rightarrow -\alpha$, $\beta\rightarrow -\beta$, the rotation matrix (36) will be unchanged. This is connected with the fact that the Cayley$-$Klein matrix (29) corresponds to the special unitary group $SU(2)$, and there is a homomorphism from SU(2) to the rotation group SO(3). Therefore, for a definite rotation matrix $R$ there are two matrices $Q$ and $(-Q)$ describing the same rotation. The group SU(2) is also isomorphic to the group of quaternions of norm $1$.

\section{The composition of two rotations}

With the help of the multiplication laws, Eq. (22), one obtains the product of two arbitrary
quaternions, $q$, $q^{\prime }$:
\begin{equation}
qq^{\prime }=\left( q_4q_4^{\prime }-q_mq_m^{\prime }\right)
e_4+\left( q_4^{\prime }q_m+q_4q_m^{\prime }+\epsilon
_{mnk}q_nq_k^{\prime }\right) e_m. \label{37}
\end{equation}
We represent the arbitrary quaternion as $q=q_4+{\bf q}$ (so $q_4e_4\rightarrow q_4$, $q_me_m\rightarrow
{\bf q}$), where $q_4$ and ${\bf q}$ are the scalar and vector
parts of the quaternion, respectively. Using these
notations, Eq. (37) can be represented as follows:
\begin{equation}
qq^{\prime }=q_4q_4^{\prime }-({\bf q}\cdot{\bf q}^{\prime
})+q_4^{\prime }{\bf q}+q_4 {\bf q}^{\prime }+ {\bf q}\times{\bf
q}^{\prime }. \label{38}
\end{equation}
Thus, the dot $({\bf q}\cdot{\bf q}^{\prime}) =q_m q_m^{\prime
}$ and cross ${\bf q}\times{\bf q}^{\prime }$ products are parts of the quaternion multiplication. It is easy to verify that the combined law of three quaternions obey the associativity law: $\left( q_1q_2\right)
q_3=q_1\left( q_2q_3\right)$ as should be for the rotation group. Replacing $q$ and $\overline{q}$ from Eqs. (25), (26) into Eq. (27), and taking into consideration the multiplication law, Eq. (37), we obtain
\begin{equation}
\textbf{x}^{\prime }=\left(\cos\alpha\right) \textbf{x} +\sin\alpha\left(\widehat{\textbf{b}}\times \textbf{x}\right)
+2\sin^2\frac{\alpha}{2}\left(\widehat{\textbf{b}}\cdot \textbf{x}\right)\widehat{\textbf{b}}.
\label{39}
\end{equation}
The same expression can be obtained from Eq. (16) and the transformation $x_i^{\prime }=R_{ij}x_j$.
Thus, two approaches based on the rotation matrix, Eq. (16), and quaternion, Eq. (25), with the transformation law, Eq. (27), are identical.

Let us consider two consecutive rotations with the quaternions $q$ and $q^{\prime}$ which are parameterized by Eq. 25. Then from the multiplication law, Eq. (37), we find the resultant rotation vector
\begin{equation}
\textbf{b}^{\prime \prime}\equiv\left(\textbf{b},\textbf{b}^{\prime}\right)=
\sqrt{1-\textbf{b}^2}\textbf{b}^{\prime}+\sqrt{1-\textbf{b}^{\prime 2}}\textbf{b}+\textbf{b}\times\textbf{b}^{\prime}.
\label{40}
\end{equation}
Squaring Eq. (40), one obtains the expression allowing us to find the resultant rotation angle
\begin{equation}
\cos\frac{\alpha^{\prime \prime}}{2}=\cos\frac{\alpha}{2}\cos\frac{\alpha^{\prime }}{2}-
\sin\frac{\alpha}{2}\sin\frac{\alpha^{\prime }}{2}\cos\beta,
\label{41}
\end{equation}
where $\beta$ is an angle between vectors $\textbf{b}$ and $\textbf{b}^{\prime }$. If $\beta=0$, i.e. vectors $\textbf{b}$ and $\textbf{b}^{\prime }$ are parallel, we obtain from Eq. (41) the trivial result $\alpha^{\prime \prime}=\alpha+\alpha^{\prime}$. One can notice that according to Eq. (40), in general, $\left(\textbf{b},\textbf{b}^{\prime}\right)\neq \left(\textbf{b}^{\prime},\textbf{b}\right)$ because the rotation group is non-commutative (non-Abelian) group. It follows from Eq. (40) that $\left(\textbf{b},\textbf{b}^{\prime}\right)^2= \left(\textbf{b}^{\prime},\textbf{b}\right)^2$, i.e. the rotation angle does not depend on the order of two successive rotations, but the direction of the rotation axis depends on the order. It should be noted that Eq. (40) gives the relation $\left(\textbf{b},-\textbf{b}\right)=0$. As a result, the inverse rotation corresponds to the parameter $-\textbf{b}$, $R(-\textbf{b})=R^{-1}(\textbf{b})$.

\section{The Rodrigues and Gibbs parameterizations}

Let us connect the modified Gibbs parametrization of the rotation matrix (8) with the Rodrigues parameters, which are defined by \cite{Rodrigues}
\begin{equation}
\textbf{r}=\hat{\textbf{b}}\alpha ,
\label{42}
\end{equation}
where $\alpha$ is an angle of the rotation in radians and $\hat{\textbf{b}}$ is a unit vector along the rotation axis. The Rodrigues parameters (Eq. (42)) possess a discontinuity at the angle of $\pi$ radians. Thus, the vector $\textbf{r}$ with the length $|\textbf{r}|=\pi$ results in the same rotation as the vector ($-\textbf{r}$). In the rotation vector space, rotations can be represented as points inside and on the surface of a sphere with the radius of $\pi$. The points at opposite ends of a diameter correspond to the same rotation and when the angle varies smoothly the rotation vector $\textbf{r}$ can jump to another end of a diameter. The difference of $\textbf{r}$ compared with $\textbf{b}$ introduced in this article is the length (normalization) of vectors: $\textbf{r}=\left(\alpha/\sin(\alpha/2)\right)\textbf{b}$. Replacing the rotation vector $\textbf{b}$ with the Rodrigues parameters $\textbf{r}$ and placing them into Eq. (40), one can obtain the composition law of two successive rotations. It should be noted that the composition laws with Euler angles and the Rodrigues parameters are cumbersome.
Another parametrization of the rotation vector was suggested by Gibbs \cite{Gibbs}, \cite{Gibbs1}
\begin{equation}
\textbf{g}= \hat{\textbf{b}}\tan\frac{\alpha}{2}.
\label{43}
\end{equation}
Taking into account Eqs. (15), (43), we obtain the connection between two vectors
\begin{equation}
\textbf{g}= \frac{\textbf{b}}{\sqrt{1-\textbf{b}^2}},~~~~\textbf{b}=\frac{\textbf{g}}{\sqrt{1+\textbf{g}^2}}.
\label{44}
\end{equation}
From Eqs. (8), (44), one finds the rotation matrix in terms of the Gibbs parameter
\begin{equation}
R_{ij}=\frac{\left(1-\textbf{g}^2\right)\delta_{ij}+2g_ig_j +2\epsilon_{ikj}g_k}{1+\textbf{g}^2}.
\label{45}
\end{equation}
The known composition law of Gibbs's parameters follows from Eqs. (40),
(43), (44), (45):
\begin{equation}
\textbf{g}^{\prime \prime}\equiv\left(\textbf{g},\textbf{g}^{\prime}\right)= \frac{\textbf{g}+\textbf{g}^{\prime}+\textbf{g}\times\textbf{g}^{\prime}}
{1-\textbf{g}\cdot\textbf{g}^{\prime}}.
\label{46}
\end{equation}
If successive rotation angles $\alpha$ and $\alpha^{\prime}$ are small, $\alpha\ll 1$, $\alpha^{\prime}\ll 1$, we obtain, from Eqs. (40), (46), the approximate relations $\textbf{b}^{\prime \prime}\approx \textbf{b}+\textbf{b}^{\prime}$, $\textbf{g}^{\prime \prime}\approx \textbf{g}+\textbf{g}^{\prime}$, i.e. the vector addition approximately holds. The disadvantage of Gibbs's parametrization follows from Eq. (43): at $\alpha=\pi$ the length of the parameter $\textbf{g}$ becomes infinity. Therefore, the direction of the vector $\textbf{g}$ is not defined at $\alpha=\pi$. At the same time the length of the parameter $\textbf{b}$ is finite, $0\leq\textbf{b}\leq 1$. It follows from Eq. (46) that if for two successive rotations $\textbf{g}\cdot\textbf{g}^{\prime}=1$, the resultant rotation corresponds to the angle $\alpha^{\prime \prime}=\pi$ because $|\textbf{g}^{\prime \prime}|=\infty$. We note that if one multiplies the rotation matrices (8) or (45) by ($-1$), then the corresponding matrix gives improper rotations including the reflection of coordinates. Thus, we have obtained the known composition law of Gibbs's parameter (46) from Eq. (40). Therefore, we suggest the simplest form of the composition law, Eq. (40), of
two successive rotations with vector $\textbf{b}=\hat{\textbf{n}}\sin(\alpha/2)$.

\section{Conclusion}

We have derived from the first principles the general expression for the rotation matrix (Eq. (8)) with non-normalized rotation vector $\textbf{b}$. This form of the rotation matrix is very convenient for practical calculations. For example, the resultant matrix for two consequent rotations (the composition of two rotations) takes a simple form shown in Eq. (40). According to the Euler theorem the direction of the vector $\textbf{b}$ defines the axis of the rotation and the length of the rotation vector is given by $|\textbf{b}|=\sin(\alpha/2)$ with $\alpha$ being the rotation angle. Eqs. (19)-(21) define the connection of the rotation vector components with Euler angles $\varphi$, $\psi$, and $\theta$. The quaternion $q$, which defines the rotation law in Eq. (27), is connected with the rotation vector $\textbf{b}$ by a simple equation (25). Cayley$-$Klein parameters $\alpha$ and $\beta$ entering the matrix $Q$, Eq. (29), are expressed via the $\textbf{b}$ by Eq. (30) and through Euler angles by Eq. (32). The rotation matrix $R$ also is conveyed via Cayley$-$Klein parameters, Eq. (36).  The Rodrigues and Gibbs representations are considered as a particular cases, and are given by Eqs. (42), (43). The composition law of two successive rotations is simpler for vector $\textbf{b}$ compared to the Gibbs vector $\textbf{g}$ or Rodrigues vector $\textbf{r}$.

The Rodrigues parameters $\textbf{r}$ with the length $|\textbf{r}|=\pi$ have the same rotation as the rotation with the vector ($-\textbf{r}$). Therefore, the Rodrigues parameters at the angle of $\pi$ possess a discontinuity. If the angle varies smoothly the rotation vector $\textbf{r}$ jumps at the angle of $\pi$ and this is the defect of Rodrigues's parametrization. In addition, the composition law for the Rodrigues parameters is complicated.
The Gibbs parametrization suffers the similar limitation. At the angle $\alpha=\pi$ the length of the parameter $\textbf{g}$ becomes infinity. As a result, the direction of the vector $\textbf{g}$ at $\alpha=\pi$ is not defined and this is the disadvantage of the Gibbs parametrization. At the same time the length of the parameter $\textbf{b}$, (see Eq. (15)), is finite.
This is the benefit of the parametrization considered.

The modified Gibbs parameterizations of the rotation matrix is more convenient to connect the rotation vector
with Euler angles, quaternion and Cayley$-$Klein parameters. The rotation using the modified Gibbs parametrization has the advantage compared to the case of other rotation methods.
The rotation matrix with the modified Gibbs parameterizations for two rotation result in two independent
equations for the vector $\textbf{b}$ (Eq. (40)) and the angle of the rotation $\alpha$ (Eq. (41)), while the vector
composition of two rotations in terms of Euler angles include a system of nine equations (see Eq. (17)).
The complexity on computing the composition of many rotations scales not as fast with the number of rotations using the modified Gibbs parameterizations representation, compared to the other methods. In the case of quaternions the composition requires the product of three quaternions to extract the rotated vector, while the modified Gibbs parameterization needs the product of two rotation matrices
 (Eq. (16)). The procedure for many angles require going back and forth to the quaternion elements to get the
resulting rotation. The relations of the modified Gibbs parameterization has simpler form for two rotations (Eq. (41)) compared
to Euler angles and Cayley$-$Klein parameterizations, which is usually accomplished via computational procedures.
Many rotations can be represented by a sequence of two rotations, therefore, Eqs. (40) and (41) will simplify the calculation procedure of many rotations. However, the increase in the number of rotations quickly proliferate the complexity of the symbolic expressions, and therefore numerical methods have to be employed for many rotations.
The modified Gibbs representation (Eq. (16)), can be applied for rotation of any rank
tensors. Therefore, it is advantageous compared to quaternions.
The method can be effectively applied in calculating nonlinear susceptibility tensors for laboratory and molecular coordinate systems in biological structures \cite{Adam}, \cite{Adam1}, \cite{Richard}.
The new approach can be employed in the fitting algorithm for orientation of cylindrical axis of nonlinear susceptibility tensor for polarimetric Stokes$-$Mueller microscopy data \cite{Samim}. Since fitting has to be accomplished for each pixel of the image,
the approach reduces substantially the calculation time.
The Euler representation of the rotation matrix by the vector $\textbf{b}$ will be applied in the future studies for calculations of orientation dependent optical responses in the field of molecular physics.


\begin{thebibliography}{99}

\bibitem{Euler} L. Euler, Novi Comment. Acad. Sci. Petrop., \textbf{20} (1775), 189-207; 208-238.


\bibitem{Gibbs} J. W. Gibbs and E. B. Wilson, Vector Analysis, New York, NY, Scribner, 1901.

\bibitem{Gibbs1} J. W. Gibbs, Scientific Papers, Vol.2, Dover Publications Inc., New York, 1961, p.65.

\bibitem{Rodrigues} O. Rodrigues, Journal de Math´ematiques \textbf{5} (1840), 380-440.

\bibitem{Euler2} L. Euler,   Novi Comment. Acad. Sci. Petrop.,  \textbf{ 15}, Section 33, (1770) 101.

\bibitem{Gauss} C. F. Gauss, Werke, Vol. VIII, pp. 357-362, G\"{o}ttingen, K\"{o}nigliche Gesellschaft der Wisssenscaften, 1900.

\bibitem{Fedorov} F. I. Fedorov, Gruppa Lorentsa, Imprint Moskva: Nauka, 1979 (in Russian).

\bibitem{Marsden} Jerrold E. Marsden, Tudor Ratiu, Introduction to Mechanics and Symmetry.
A Basic Exposition of Classical Mechanical Systems, Springer, 1999.

\bibitem{Joshi}	A. W. Joshi, Elements of Group Theory for Physicists, New York: Wiley, 1982.



\bibitem{Goldstein} H. Goldstein, Classical Mechanics, 2$^{nd}$ ed., Reading, MA: Addison-Wesley, 1980.

\bibitem{Landau} L. D. Landau and E. M. Lifschitz, Mechanics, Oxford, England: Pergamon Press, 1976.

\bibitem{Casanova} G. Casanova, L'alg\`ebre Vectorielle, Presses Universitaires de France, 1976.

\bibitem{Wertz} J. R. Wertz, ed., Spacecraft Attitude Determination and Control, Dordrecht, Holland, D. Reidel, 1978.

\bibitem{Altmann} S. L. Altmann, Mathematics Magazine \textbf{62}, No. 5 (1989), 291-3138; Rotations, quaternions, and double groups,
Clarendon Press, Oxford, 1986.

\bibitem{Shuster} M. D. Shuster, Journal of the Astronautical Sciences \textbf{41}, No. 4
(1993), 439-517.

\bibitem{Kruglov2} S. I. Kruglov, Symmetry and electromagnetic interaction of fields with multi-spin.
A Volume in Contemporary Fundamental Physics, New York: Nova Science Publishers, Huntington, 2001.

\bibitem{Cayley} A. Cayley, Cambridge Mathematical Journal, \textbf{3}
(1843), 141-145; 224-232, also in The Collected Mathematical Papers of Arthur Cayley, Vol. I,
The Cambridge University Press, 1889. Johnson Reprint Corp. New York, 1963, pp. 28-35; pp. 123-126.

\bibitem{Arfken} G. Arfken, Mathematical Methods for Physicists, New York: Academic Press, 1985.

\bibitem{Gel'fand} I. M. Gel'fand, R. A. Minlos and Z. Ya. Shapiro, Representations
of the Rotation and Lorentz Groups and their Applications, Pergamon, New York, 1963.

\bibitem{Hamilton} W. R. Hamilton,  Philosophical Magazine, 3$^{nd}$ Series, \textbf{25} (1844), 489-495.

\bibitem{Bar} I. Y. Bar-Itzhack and Y. Oshman, IEEE Transactions on Aerospace and Electronic Systems,
\textbf{AES-21}, No.1 (1985), 128-135.

\bibitem{Markley} F. L. Markley, Journal of Guidance, Control and Dynamics \textbf{26}, No.2 (2003), 311-317; \textbf{31}, No.2 (2008), 440-442.

\bibitem{Markley1} F. L. Markley, Journal of the Astronautical Sciences \textbf{52}, Nos. 1-2, (2004), 221-238; \textbf{54}, Nos. 3–4 (2006), 391–413.

\bibitem{Kruglov} S. I. Kruglov,  Annales Fond. Broglie \textbf{27} (2002), 343-358
[arXiv:hep-th/0110059].

\bibitem{Kruglov1} S. I. Kruglov, Int. J. Theor. Phys. \textbf{41} (2002), 653-687 [arXiv:hep-th/0110251].

\bibitem{Adam} A. E. Tuer, S. Krouglov, N. Prent, R. Cisek, D. Sandkuijl, K. Yasufuku, B. C. Wilson, and V. Barzda, J. Phys. Chem. B \textbf{115} (2011), 12759-12769.

\bibitem{Adam1} A. E. Tuer, S. Krouglov, R. Cisek, D. Tokarz, and V. Barzda, J. Comp. Chem., \textbf{32} (2011), 1128-1134.

\bibitem{Richard} D. Tokarz, R. Cisek, S. Krouglov, L. Kontenis, U. Fekl, and V. Barzda, J. Phys. Chem. B \textbf{118} (2014), 3814-3822.

\bibitem{Samim} M. Samim, S. Krouglov, D. F. James, and V. Barzda, J. Opt. Soc. Am. B, \textbf{33} (2016), 2617-2625.

\end{thebibliography}
\end{document}